\documentclass[aip]{revtex4-1}
\usepackage{graphicx}
\usepackage{dcolumn}
\usepackage{bm}

\usepackage[utf8]{inputenc}
\usepackage[T1]{fontenc}
\usepackage{mathptmx}
\usepackage{etoolbox}
\usepackage{float} 

\draft 
\makeatletter
\def\@email#1#2{%
 \endgroup
 \patchcmd{\titleblock@produce}
  {\frontmatter@RRAPformat}
  {\frontmatter@RRAPformat{\produce@RRAP{*#1\href{mailto:#2}{#2}}}\frontmatter@RRAPformat}
  {}{}
}%
\makeatother
\begin{document}


\title{High yield, low disorder Si/SiGe heterostructures for spin qubit devices manufactured in a BiCMOS pilot line} 




\author{Alberto Mistroni}
 \altaffiliation[contact: ]{mistroni@ihp-microelectronics.com}
\author{Marco Lisker}
\author{Yuji Yamamoto}
\author{Wei-Chen Wen}
\author{Fabian Fidorra}
\author{Henriette Tetzner}
\affiliation{ IHP - Leibniz Institute for High Performance Microelectronics, Frankfurt (Oder) 15236, Germany}

\author{Laura K. Diebel}
\affiliation{Fakultät für Physik, Universität Regensburg, Regensburg 93040, Germany
}

\author{Lino Visser}
\affiliation{JARA-FIT Institute for Quantum Information, Forschungszentrum Jülich GmbH and RWTH Aachen University, 52062 Aachen}
\author{Spandan Anupam}
\affiliation{JARA-FIT Institute for Quantum Information, Forschungszentrum Jülich GmbH and RWTH Aachen University, 52062 Aachen}
\author{Vincent Mourik}
\affiliation{JARA-FIT Institute for Quantum Information, Forschungszentrum Jülich GmbH and RWTH Aachen University, 52062 Aachen}
\author{Lars R. Schreiber}
\affiliation{JARA-FIT Institute for Quantum Information, Forschungszentrum Jülich GmbH and RWTH Aachen University, 52062 Aachen}
\affiliation{ARQUE Systems GmbH, 52074 Aachen, Germany}
\author{Hendrik Bluhm}
\affiliation{JARA-FIT Institute for Quantum Information, Forschungszentrum Jülich GmbH and RWTH Aachen University, 52062 Aachen}
\affiliation{ARQUE Systems GmbH, 52074 Aachen, Germany}

\author{Dominique Bougeard}
\affiliation{Fakultät für Physik, Universität Regensburg, Regensburg 93040, Germany
}%

\author{Marvin H. Zoellner}
\affiliation{ IHP - Leibniz Institute for High Performance Microelectronics, Frankfurt (Oder) 15236, Germany}
\author{Giovanni Capellini}
\affiliation{ IHP - Leibniz Institute for High Performance Microelectronics, Frankfurt (Oder) 15236, Germany}
\affiliation{Dipartimento di Scienze, Università Roma Tre, Roma 00146, Italy.}

\author{Felix Reichmann}
\altaffiliation[contact: ]{reichmann@ihp-microelectronics.com}
\affiliation{ IHP - Leibniz Institute for High Performance Microelectronics, Frankfurt (Oder) 15236, Germany}

\date{\today}

\begin{abstract}

The prospect of achieving fault-tolerant quantum computing with semiconductor spin qubits in Si/SiGe heterostructures relies on the integration of a large number of identical devices, a feat achievable through a scalable (Bi)CMOS manufacturing approach. To this end, both the gate stack and the Si/SiGe heterostructure must be of high quality, exhibiting uniformity across the wafer and consistent performance across multiple fabrication runs. Here, we report a comprehensive investigation of Si/SiGe heterostructures and gate stacks, fabricated in an industry-standard 200 mm BiCMOS pilot line. We evaluate the homogeneity and reproducibility by probing the properties of the two-dimensional electron gas (2DEG) in the shallow silicon quantum well through magnetotransport characterization of Hall bar–shaped field-effect transistors at 1.5 K. Across all the probed wafers, we observe minimal variation of the 2DEG properties, with an average maximum mobility of $(4.25\pm0.17)\times 10^{5}$ cm$^{2}$/Vs and low percolation carrier density of $(5.9\pm0.18)\times 10^{10}$ cm$^{-2}$ evidencing low disorder potential in the quantum well. The observed narrow statistical distribution of the transport properties highlights the reproducibility and the stability of the fabrication process. Furthermore, wafer-scale characterization of a selected individual wafer evidenced the homogeneity of the device performances across the wafer area. Based on these findings, we conclude that our material and processes provide a suitable platform for the development of scalable, Si/SiGe-based quantum devices.
\end{abstract}


\maketitle 

Spin qubits based on gate-defined quantum dots in Si hold great promise as scalable building blocks for fault-tolerant quantum computers \cite{SiQubitsReview}, leveraging on the mature complementary metal-oxide-semiconductor (CMOS) technology \cite{Intel300}\cite{Intel12qbit}\cite{QDArray}. The quantum dots are electrostatically defined by applying biases to nanostructured gates, which locally modulate the two-dimensional charge density and enable control over individual charge and spin states. Shallow silicon quantum wells (Si QW) in undoped Si/SiGe heterostructures have been developed to reduce the charge noise from defects at the dielectric interface, thereby increasing the qubit coherence times compared to qubits based on the Si-MOS platform  \cite{ReviewSpinQubits}\cite{HighFidCH}\cite{HighFid}\cite{Coherence}\cite{Struck2020_longT2}. Using Si/SiGe-based devices, significant milestones towards scalable quantum computing have already been achieved, including the demonstration of a 6 qubit quantum processor \cite{6QubitProc} and 12 qubit arrays fabricated in a industrial manufacturing lines \cite{Intel12qbit} \cite{Huck2025_industrialQubit}. 
Furthermore, the Si/SiGe platform has been used to demonstrate charge-shuttling lines to connect distant qubit arrays, as well as alternative architectures that help overcome one of the main challenges to scalable spin-based quantum computing which is the wiring congestion.\cite{SparseQubitsArray}\cite{SpinBus}\cite{Shuttling2}\cite{RUOYU2018}. Despite recent advancements, the performance of quantum dots and qubits remains sensitive to the quality of both the gate stack and the underlying heterostructure, making material and process improvements essential \cite{MaterialReview}. Furthermore, the scaling to large numbers of devices requires process reproducibility and homogeneous performance across large wafer areas, underscoring the need for large-scale fabrication and testing \cite{Intel300}. Hall bar field-effect transistors (HB-FETs) are test structures that can be fabricated on SiGe heterostructures using gate stacks similar to those in quantum dot and qubit devices. Through magnetotransport measurements, they enable the characterization of key transport properties of the two-dimensional electron gas
(2DEG) confined in the Si quantum well that cannot be directly measured in quantum dot devices, providing insights into both heterostructure and gate stack quality \cite{DegliEsposti2024}\cite{Diebel2025_BiasCool}. Compared to quantum dots and qubits, HB-FETs are simpler to fabricate and characterize, making them well-suited for evaluating wafer-scale uniformity and process stability with respect to the field-effect stack \cite{Multiplex}.

In this letter, we report a wafer-to-wafer and wafer-scale characterization of the transport properties of a Si/SiGe qubit-compatible field-effect stack, using HB-FETs fabricated in an industry-standard 200 mm BiCMOS pilot line. Our results demonstrate a reproducible process for fabricating low disorder Si/SiGe heterostructures and gate stacks, showing minimal variability across devices and wafer runs.\\


\begin{figure}[ht]
\centering
\includegraphics[width=0.5\linewidth]{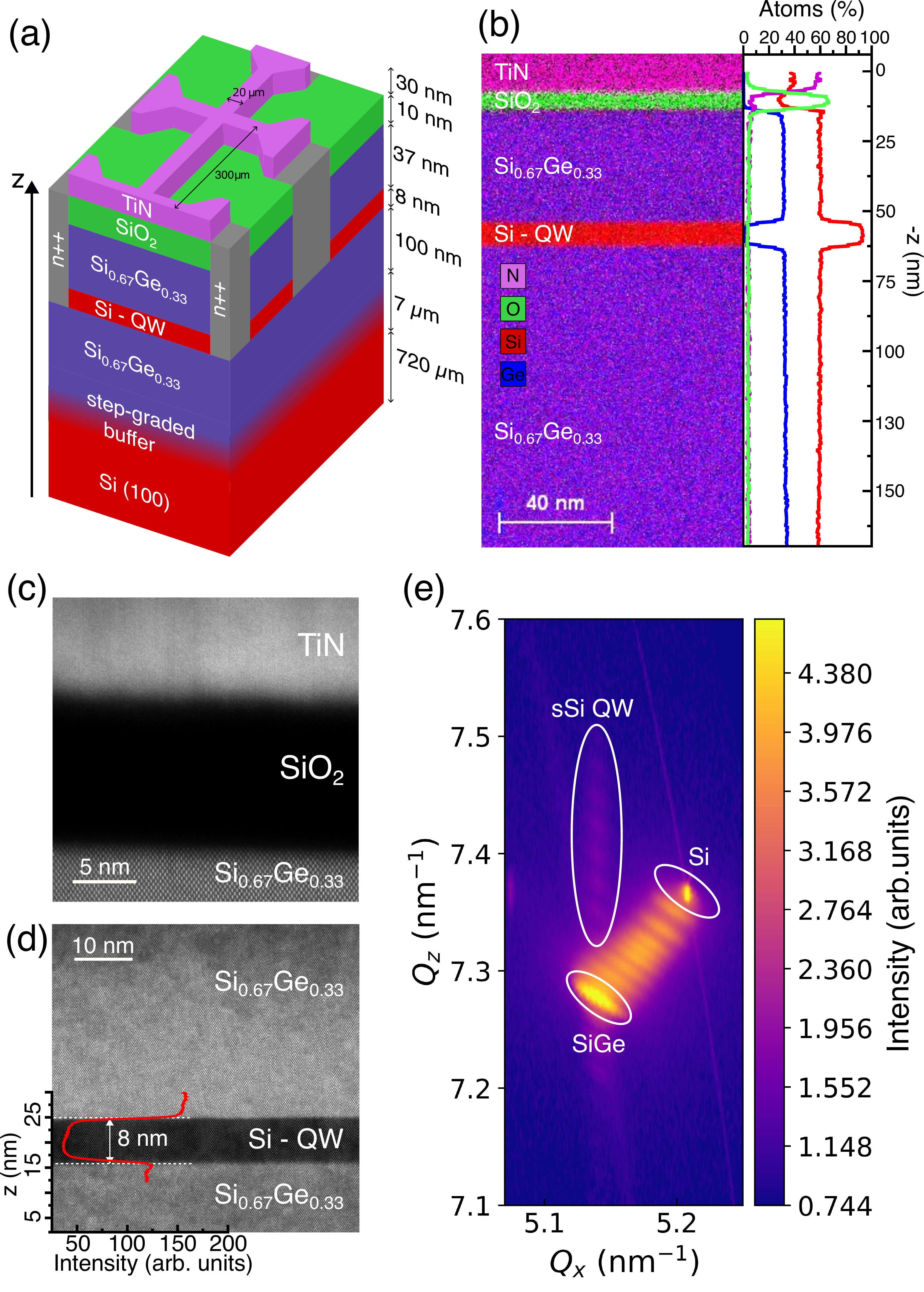}
\caption{(a) Schematic of the Si/SiGe heterostructure with fabricated HB-FET device, including the layer and device dimensions. (b) Scanning transmission electron microscopy with energy dispersive X-ray (STEM/EDX) of the device layer stack down to the bottom SiGe barrier with line-cut profiles of the Si (red), Ge (blue), O (green) and N (magenta) concentrations on the right side. (c) High-angle annular dark-field (HADDF) STEM image of the SiGe/oxide interface.  (d) HADDF STEM cross section of the Si quantum well interface with superimposed intensity profile (red line) used to estimate effective QW thickness. (e) X-ray diffraction reciprocal space map around (422) reflections from the Si substrate, the SiGe buffer and the Si QW marked with white labels and circular symbols.}
\label{Fig:StructChar}
\end{figure}

All Si/SiGe heterostructures reported within here were grown using industry-standard reduced-pressure chemical vapor deposition (RP-CVD) with silane and germane precursors \cite{Felix}, making the process readily transferable to isotopically enriched alternatives required for nuclear spin-free heterostacks, that enable high qubit coherence times \cite{Coherence}\cite{Struck2020_longT2}\cite{Domonique_old1}\cite{Domonique_old2}. Fig. \ref{Fig:StructChar}a shows a schematic of the grown heterostructure and the fabricated HB-FETs. Starting from 200 mm p-doped (5–22 $\Omega$cm) Si (100) wafers, a pre-epi clean of the wafer was performed, followed by the growth of a 4 $\mu$m thick step-graded Si$_{1-x}$Ge$_x$  virtual substrate,  terminating with a nominal Ge concentration of $x$= 0.33 $\pm$ 0.01. On top of the graded virtual substrate, a 2.7 $\mu$m thick constant composition Si$_{0.67}$Ge$_{0.33}$ buffer layer was first deposited and then chemical mechanical polishing was applied to reduce the surface roughening caused by the cross-hatch pattern formation \cite{GiovanniCHP}. After a further surface clean and the growth of a 100 nm-thick Si$_{0.67}$Ge$_{0.33}$ bottom barrier, an 8 nm Si QW was grown, followed by a 37 nm Si$_{0.67}$Ge$_{0.33}$ top barrier, all at 600 \textdegree C. The heterostructure was terminated with a 6 nm-thick epitaxial, sacrificial Si cap layer grown at 700 \textdegree C, which protects the underlying SiGe barrier from native oxidation \cite{SelectiveOxidationLiou}\cite{SelectiveOxidationLeGoues}. Following heterostructure growth, HB-FETs were fabricated using CMOS compatible processes, described in more detail here \cite{Felix}. Ohmic contacts were formed by selective phosphorus ion implantation with a dose of $4.5\times 10^{15}$ cm$^{-2}$ and an acceleration energy of 20 keV, followed by one minute thermal annealing at 700 \textdegree C to activate the implanted phosphorus. A high-density plasma (HDP) SiO$_2$ layer with 10 nm thickness was deposited at a temperature of 300 \textdegree C as the gate dielectric. The Si cap was gradually consumed during fabrication, due to repeated formation of (native) SiO$_2$ and subsequent HF dips. It was then completely consumed after the HDP SiO$_2$ deposition, as confirmed by in-line ellipsometry measurements. A TiN top gate layer was then deposited by physical vapor deposition (PVD) and structured to Hall bar shape, using optical lithography and reactive ion etching. The fabricated Hall bars feature a channel length of 2 mm, a channel width of 20 $\mu$m and pitch between two cross points of 300 $\mu$m.\\

The structural characteristics of the field-effect stack post device fabrication were investigated using energy-dispersive X-ray spectroscopy (EDX), scanning transmission electron microscopy (STEM), and X-ray diffraction (XRD). Fig. \ref{Fig:StructChar}b shows a STEM/EDX cross-section view of the active region, illustrating the well-defined Si quantum well, the surrounding SiGe barriers, the SiO$_2$ dielectric and the TiN metal layer. The corresponding chemical composition depth profile on the right side of the figure confirms that the Ge concentration in the SiGe layers is close to the target value of 33\%.
High-angle annular dark-field STEM (HAADF-STEM) was used to examine the SiGe/SiO$_{2}$ interface and Si quantum well region of the fabricated device. The cross-section shown in Fig. \ref{Fig:StructChar}c confirms the absence of any residual Si cap and reveals no significant Ge pile-up, which has previously been observed in similar heterostructures featuring a non-epitaxial Si cap and native oxide formation \cite{Dingle}. In contrast, our observations indicate that the SiGe top barrier remained protected from native oxidation during the fabrication process. Fig. \ref{Fig:StructChar}d shows the cross-sectional HAADF STEM image of the Si QW region. To quantitatively assess the QW interfaces sharpness, a sigmoid fit was applied to the STEM intensity profile, from which the interface width was extracted as 4$\tau$ parameter \cite{InterfaceWEst}\cite{Klos2024_4tau}. We obtain a top SiGe/Si QW interface width of 4$\tau_{top}$= 0.79 nm and a Si QW/bottom SiGe interface width of $\tau_{bot}$=1.04 nm, values comparable with state-of-the-art CVD-grown Si/SiGe heterostructures, used for quantum device fabrication \cite{DegliEsposti2024}. The intensity profile was also used to determine the effective quantum well thickness, which was found to be around 8 nm.
The in-plane tensile strain in the Si QW, induced by pseudomorphic growth on the nominally relaxed 100 nm Si$_{0.67}$Ge$_{0.33}$  buffer layer, was quantified by X-ray diffraction (XRD) in a wafer region adjacent to a representative device. Fig. \ref{Fig:StructChar}e presents the reciprocal space map around (422) reflections with features corresponding to the Si (100) substrate, SiGe buffer and the Si QW highlighted by white circular shapes. The diffraction signal arising from the step-graded region of the virtual substrate is clearly visible between the Si$_{0.67}$Ge$_{0.33}$(422) and the Si(422) reciprocal lattice points. The SiGe buffer exhibits a negligible residual strain ($\varepsilon_{||}^{\text{SiGe}}$=0.04\%), attributed to the difference in thermal expansion of this layer compared to that of the Si substrate, while the Si QW is pseudomorphic with a strain of $\varepsilon_{||}^{\text{Si}}$=1.29$\pm$ 0,01 \%, in agreement with expectations. \\


A systematic magnetotransport characterization of Si/SiGe HB-FETs was performed in order to assess the average performance and variability of key transport properties of the 2DEG across multiple wafers and multiple devices. For this purpose, up to four devices were bonded to an LCC-44 chip carrier and loaded into an Oxford Instruments Teslatron-PT cryostat. Once the base temperature of 1.5 K was reached, the 2DEG in the Si QWs was first accumulated by applying to the gate electrode a DC gate voltage $V_G$ larger than the threshold voltage V$_{\text{thr}}$. Quantum Hall-effect measurements were conducted by sweeping the magnetic field  $B_{\bot}$ perpendicular to the sample surface,  while measuring the longitudinal ($V_{xx}$) and transverse ($V_{xy}$) voltages, along with the channel current using a four-probe lock-in technique. From the calculated longitudinal ($\rho_{xx}$) and transverse ($\rho_{xy}$) resistivities, we extracted the density and mobility values inverting the relations $\rho_{xy} = B_{\bot}/en$, and $\sigma_{xx} = ne\mu$, where $e$ is the elemental charge and $\sigma_{xx}$ is the transversal conductivity obtained inverting the resistivity tensor. To ensure the validity of the findings, two devices were measured in independent laboratories (see supplementary information), showing consistent results with those reported in this work. To evaluate the consistency and reproducibility of device performance across wafers from different fabrication runs, 6 nominally identical wafers were grown (wafers A to F), with one sample consistently selected from the center area of each wafer for characterization. In addition, the uniformity of performance across the 200 mm wafer area was evaluated by characterizing 12 additional devices from wafer A.

\begin{figure}[htbp!]
\centering
\includegraphics[width=1\linewidth]{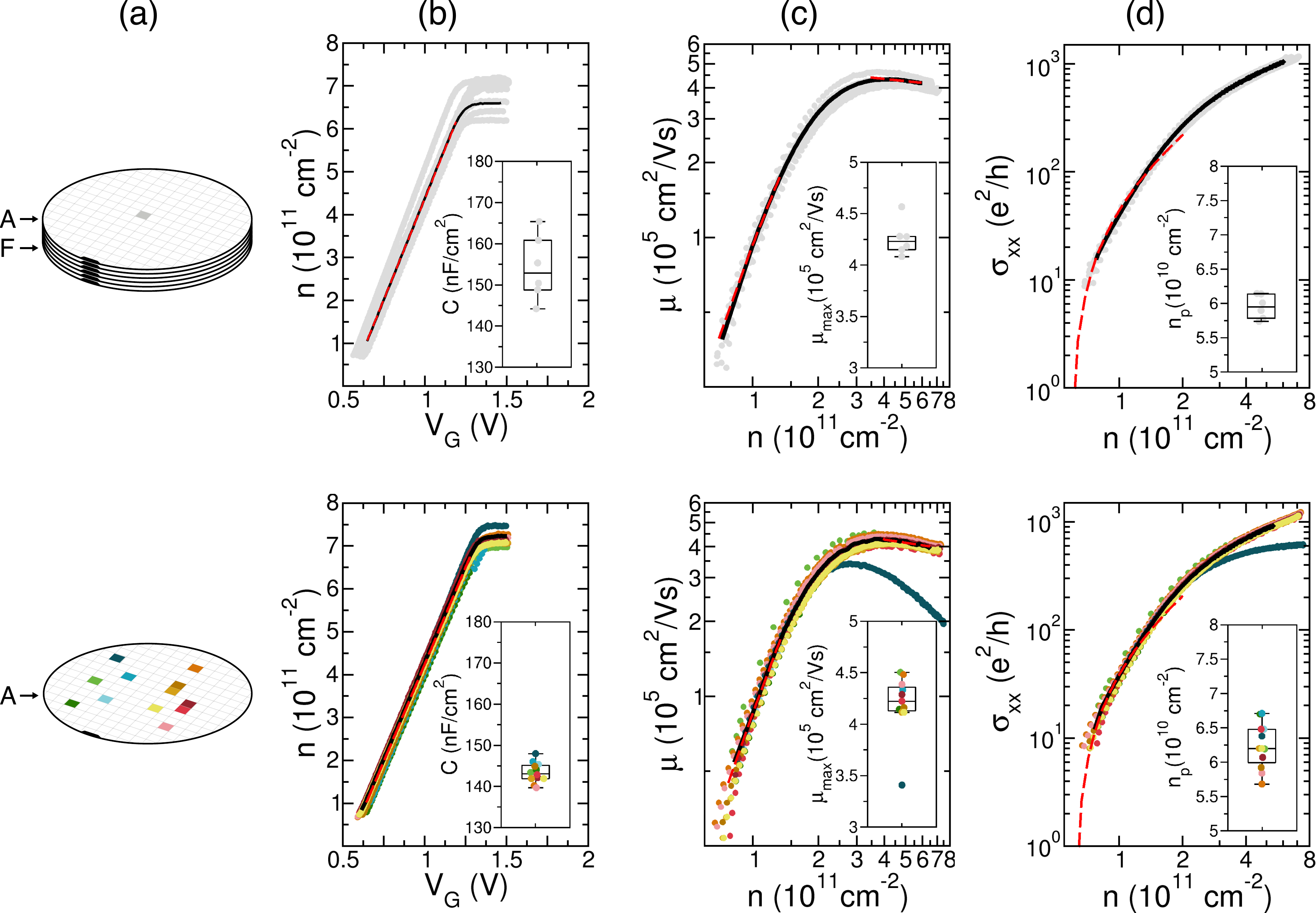}
\caption{Magnetotransport characterization at 1.5 K. The upper panel corresponds to the wafer-to-wafer characterization (wafers A to F, 6 devices in total) while the lower panel to the wafer-scale characterization of a single wafer (wafer A, 12 devices in total). (a) Wafers dicing scheme with colored squares denoting the position of the devices selected for the wafer-to-wafer (grey) and wafer-scale (different color codes) transport characterization. (b) Carrier density $n$ plotted as a function of gate voltage $V_G$. The red dashed line represents the linear fit to the mean trend shown by the black solid line. (c) Hall mobility $\mu$  plotted as a function of $n$ on a double-logarithmic scale. The mean trend (black solid) is fitted with a power-law dependence (red dashed). (d) Double-logarithmic plot of longitudinal conductivity $\sigma_{xx}$ against $n$, with the percolation density $n_p$ extracted from fits (red dashed) to the mean curve (black solid). Boxplots in the insets show the distributions of capacitance, peak Hall mobility and extracted $n_p$ values, respectively.}
\label{Fig:ClassicalTransport}
\end{figure}

The main results of the magnetotransport characterization are presented in Fig. \ref{Fig:ClassicalTransport}, where the top panel compares the performance of individual devices from wafers A - F, and the bottom panel shows a comparison among 12 additional devices from wafer A. The colored squares highlighted in \ref{Fig:ClassicalTransport}a indicate the locations on the wafer of the devices selected for characterization. Fig. \ref{Fig:ClassicalTransport}b illustrates the dependence of the 2DEG carrier density ($n$) in the Si QW on the gate voltage ($V_{\text{G}}$) for the measured HB-FETs.  For $V_{\text{G}}$ > $V_{\text{thr}}$, the electron density $n$ increases linearly with $V_{\text{G}}$. This behavior is consistent with the parallel-plate capacitor model, where the top SiGe barrier and the oxide layer act as dielectrics between the 2DEG and the TiN metal layer. At higher gate voltages the additional electrons injected into the QW from the ohmic contacts tunnel through the top SiGe barrier to the surface and get trapped by defect states at the dielectric interface \cite{DensSat}, leading to a carrier density plateau at around $7\times 10^{11}$ cm$^{-2}$.We note in Fig. \ref{Fig:ClassicalTransport}b that the carrier density saturation value varies across wafers fabricated in different runs. This variation can be attributed to differences in the quantum well thickness and/or in the Ge concentration of the top barrier, both of which affect the electronic confinement in the quantum well \cite{DensSat}. Nevertheless, this variation has no significant impact on the transport performance, as shown in Fig. \ref{Fig:ClassicalTransport}c and d. By applying a linear fit in the $V_{\text{G}}$ range of 0.6V – 1.2V  we estimate the capacitance per unit area using C = e d$n$/d$V_{\text{G}}$, which can be used to evaluate the thickness of the dielectric stack. We validate the accuracy of this method by comparing the fitted capacitance of a selected device with the effective capacitance calculated as $C_{\text{eff}}$ = (1/$C_{\text{Ox}}$ + 1/$C_{\text{SiGe}}$)$^{-1}$ using top barrier and oxide thickness values extrapolated from a STEM image of an area nearby the measured HB-FET, and $\varepsilon_{\text{Ox}} = 3.9$ and $\varepsilon_{\text{SiGe}} = 13.13$ as dielectric constant for the oxide layer and the SiGe barrier, respectively \cite{Capacitance}. We found that the capacitance derived by the $n(V_{\text{G}})$ slope agrees within 1\% with the calculated effective capacitance, demonstrating consistency between both methods.  From the statistical characterization reported in the insets of Fig. \ref{Fig:ClassicalTransport}b we observe only small wafer-to-wafer and wafer-scale capacitance variations of 5\% and 1.7\%,  respectively. These results highlight the SiGe and oxide thickness uniformity across the wafer area and the reproducibility of the deposition and fabrication processes over multiple wafer runs.

Fig. \ref{Fig:ClassicalTransport}c presents the density-dependent mobility ($\mu$) plots of all the measured HB-FETs. The mobility behavior is similar for most of the probed devices: it increases sharply in the low-density regime ($n < 2\times 10^{11}$ cm$^{-2}$), peaks around $n \sim 3\times 10^{11}$ cm$^{-2}$ and then slightly decreases in the high-density regime above $4\times 10^{11}$ cm$^{-2}$. The fit of $n(\mu)$ with the power-law relation $n \sim \mu^{\alpha}$ in the low-density region leads to a median value $\alpha$= 2.47. If if we consider local-field corrections to the theoretical calculations, this indicates Coulomb scattering from remote impurities \cite{Laroche}, likely located at semiconductor/dielectric interface. As expected, at higher electron densities, self-screening of charges significantly reduces the impact of Coulomb scattering in the 2DEG resulting in scattering predominantly from sources near or within the quantum well, as confirmed by a decrease in the median power-law exponent $\alpha$ to -0.12. The maximum mobility distribution from the wafer-to-wafer characterization displayed in top panel of Fig. \ref{Fig:ClassicalTransport}c shows a remarkably high average value of ($4.25\pm0.17$)$\times 10^{5}$ cm$^{2}$/Vs  with a maximum measured value exceeding $4.5\times 10^{5}$ cm$^{2}$/Vs, showcasing a substantial improvement to our previous work \cite{Felix} and current state-of-the-art industrially manufactured Si/SiGe field-effect stack performances \cite{Multiplex}\cite{Intel300}. This result is corroborated by the wafer-scale characterization evidencing a similar maximum mobility average of ($4.20\pm0.28$)$\times 10^{5}$ cm$^{2}$/Vs. Only one device located near the wafer edge (depicted in dark blue in Fig. \ref{Fig:ClassicalTransport}a) shows a significantly reduced peak mobility and a high-density alpha value of -0.6,  i.e. far outside the standard deviation ranges, potentially indicating local increase of interface roughness, threading dislocations density, and/or impurities inside the QW \cite{Monroe}\cite{DasSarma}. Another possible explanation is intersubband scattering between a bound state and a strongly level-broadened continuum resonance, mediated by Coulomb impurities as shown in \cite{Huang2024_interssuband}. Despite that, its behavior in the low-density region is similar to that of all the other probed devices, thus it still ensures a low-disorder environment in the density regime where quantum dots devices are typically operated. 

We further evaluate QW potential disorder at lower electron densities by calculating the critical density $n_p$ required to achieve metallic conduction in the channel. The low-density conductivity data was fitted to the percolation model $\sigma_{xx} \propto (n-n_p)^p$, where $p=1.31$ is the critical exponent for a 2D system \cite{Percolation}. To guarantee a correct data comparison we consistently used the same electron density fitting range ($n_{min} \leq n_{range} \leq 1.2 \times 10^{11}$ cm$^{-2}$) to better capture the conductance behavior in the low density region and extrapolate it to lower densities. As shown in Fig. \ref{Fig:ClassicalTransport}d, both wafer-to-wafer and wafer-scale characterization show low average percolation values of ($5.9\pm0.18$)$\times 10^{10}$ cm$^{-2}$ and ($6.24\pm0.33$)$\times 10^{10}$ cm$^{-2}$, respectively. The limited standard deviations, below 6\% and 3\%, evidence once more the highly homogeneous minimal disorder potential landscape in the channel and is comparable with state-of-the-art field effect stacks used for quantum device fabrication \cite{Multiplex}\cite{DegliEsposti2024}. Importantly, we also observe no correlation between low $n_p$ and high $\mu_{max}$ values, underlining that different scattering mechanisms dominate at low and high densities, respectively.\\


In conclusion, we demonstrate scalable, high-quality field-effect stacks based on Si/SiGe heterostructures, fabricated using industry-compatible processes in a 200 mm BiCMOS pilot line. Building on previous work on HB-FET devices fabricated on Si quantum well heterostructures \cite{Felix}\cite{Diebel2025_BiasCool}, we report significant improvements in overall 2DEG transport properties, along with consistent and reproducible device performance across the wafer and over multiple fabrication cycles. The high mean peak mobility and low percolation density, along with their narrow distributions extracted from magnetotransport measurements at 1.5 K, are comparable to or exceed those of state-of-the-art field-effect stacks currently used in spin qubit fabrication \cite{Intel300}\cite{Intel12qbit}\cite{DegliEsposti2024}\cite{Multiplex}\cite{Dingle}. 


\section*{Supplementary Material}
See the supplementary material for the QW interface width estimation, the effective capacitance calculation and the comparison of magnetotransport results obtained in different laboratories.

\begin{acknowledgments}
This work is part of the joint project QUASAR „Halbleiter-Quantenprozessor mit shuttlingbasierter skalierbarer Architektur“ and is supported by the German Federal Ministry of Research, Technology and Space. We acknowledge the financial support of the Deutsche Forschungsgemeinschaft (DFG, German Research Foundation) via Project-ID No. 289786932 (BO 3140/4-2) and the support of the Bavarian state government with funds from the Hightech Agenda Bavaria, via the Munich Quantum Valley (MQV). Part of this work has been carried out within the Joint Lab “Spin-Based Quantum Computing” established between IHP - Leibniz Institute for High Performance Microelectronics, RWTH Aachen University and  Forschungszentrum Jülich. 

The following article has been submitted to Applied Physics Letters.
Copyright (2025) Author(s). This article is distributed under a Creative Commons Attribution-NonCommercial 4.0 International (CC BY-NC-SA) License.

\end{acknowledgments}

\section*{Conflict of Interest}
The authors have no conflicts to disclose.

\section*{Author contributions }
\textbf{A. M}.:  Investigation, Formal analysis, Visualization, Writing – original draft; \textbf{M. L.}, \textbf{Y. Y.}, \textbf{W.-C. W.},\textbf{ M. H. Z.}: Writing – review \& editing, Resources; \textbf{L. K. D.},\textbf{L. V.},\textbf{S. A.}: Writing – review \& editing, Investigation, Validation; \textbf{F. F.}, \textbf{H. T.}, \textbf{G. C. }, \textbf{V. M. }, \textbf{ L. R. S.}: Writing – review \& editing; \textbf{D. B.}, \textbf{H. B.}: Writing – review \& editing, Funding acquisition;
\textbf{F. R.:} Writing – review \& editing, Funding acquisition, Conceptualization, Project administration

\section*{Data Availability Statement}

The data that support the ﬁndings of this study are available
from the corresponding authors A.M. and F.R. upon request.


%
%

%


\bibliography{MainBib}

\end{document}